\documentclass[conference,letterpaper]{IEEEtran}
\IEEEoverridecommandlockouts
\usepackage[T1]{fontenc}
\usepackage{amsmath,amssymb,amsfonts}
\usepackage{algorithm,algorithmic}
\usepackage{float}
\usepackage{multirow}  
\usepackage{array}
\usepackage[caption=false,font=footnotesize,labelfont=rm,textfont=rm]{subfig}
\usepackage{stfloats}
\usepackage{url}
\usepackage{verbatim}
\usepackage{graphicx}
\usepackage{xcolor}
\usepackage{balance}
\usepackage[numbers,sort&compress]{natbib}
\usepackage{newtxtext}
\usepackage{newtxmath}
\usepackage[none]{hyphenat}
\usepackage{courier}
\newcommand{\figref}[1]{Fig.~\ref{#1}}
\newcommand{\subfigref}[2]{Fig.~\ref{#1}\subref{#2}}
\newcommand{\subfigrefcon}[3]{Fig.~\ref{#1}\subref{#2}$\sim $\subref{#3}}
\newcommand{\tble}[1]{TABLE~\ref{#1}}
\newcommand{\eqt}[1]{(\ref{#1})}
\def\BibTeX{{\rm B\kern-.05em{\sc i\kern-.025em b}\kern-.08em
    T\kern-.1667em\lower.7ex\hbox{E}\kern-.125emX}}
\begin{document}

\title{Dual-Domain Constraints: Designing Covert and Efficient Adversarial Examples for Secure Communication\\
\thanks{This work of Tailai Wen, Da Ke, Xiang Wang and Zhitao Huang was supported by the Natural Science Foundation of China (Grant No. 62271494).(Corresponding author: Xiang Wang).}
\thanks{This work of Tailai Wen has been supported by the Postgraduate Research Innovation Programme under Grant No. XJZH2024036.}
}

\author{\IEEEauthorblockN{1\textsuperscript{st} Tailai Wen}
\IEEEauthorblockA{\textit{College of Electronic Science and Technology} \\
\textit{National University of Defense Technology}\\
Changsha, China \\
wentailai@nudt.edu.cn}\\

\IEEEauthorblockN{3\textsuperscript{rd} Xiang Wang}
\IEEEauthorblockA{\textit{College of Electronic Science and Technology} \\
\textit{National University of Defense Technology}\\
Changsha, China \\
christopherwx@163.com} 
\and
\IEEEauthorblockN{2\textsuperscript{nd} Da Ke}
\IEEEauthorblockA{\textit{College of Electronic Engineering} \\
\textit{National University of Defense Technology}\\
Hefei, China \\
1747884404@qq.com} \\

\IEEEauthorblockN{4\textsuperscript{th} Zhitao Huang}
\IEEEauthorblockA{\textit{College of Electronic Engineering} \\
\textit{National University of Defense Technology}\\
Hefei, China \\
huangzhitao@nudt.edu.cn} 
}

\maketitle

\begin{abstract}
The advancements in Automatic Modulation Classification (AMC) have propelled the development of signal sensing and identification technologies in non-cooperative communication scenarios but also enable eavesdroppers to effectively intercept user signals in wireless communication environments. 
To protect user privacy in communication links, we have optimized the adversarial example generation model and introduced a novel framework for generating adversarial perturbations for transmitted signals. 
This framework implements dual-domain constraints in both the time and frequency domains, ensuring that the adversarial perturbation cannot be filtered out. 
Comparative experiments confirm the superiority of the proposed method and the concealment of the adversarial examples it generates.
\end{abstract}


\section{Introduction}
Automatic Modulation Classification (AMC) aims to autonomously detect and identify the modulation types of communication signals \cite{DCariticlerf1}. 
This technology addresses challenges such as spectrum scarcity and increasingly complex signal environments caused by the surge in communication devices, and has been widely applied in spectrum management for Internet of Things (IoT) devices, military signal reconnaissance, and the detection of illegal signal sources in public communication environments \cite{DCariticlerf2}. 
Recent research shows that Deep Neural Network (DNN)-driven AMC methods significantly outperform traditional statistical approaches in low Signal-to-Noise Ratio (SNR) and unknown signal scenarios \cite{DCariticlerf3}. 
However, this technology also provides eavesdroppers with more efficient means to intercept user signals \cite{DCariticlerf4}, posing significant challenges to the security of wireless communication links.

DNN-driven AMC can lead to severe misclassification when signals are subjected to carefully crafted small perturbations, known as adversarial perturbations, with the modified signals referred to as adversarial examples \cite{DCariticlerf5}. 
The transmitter can evade the eavesdropper's AMC detection by utilizing adversarial perturbations on the transmitted signal, thereby protecting the communication link \cite{DCariticlerf6}. 
Methods such as Fast Gradient Sign Method (FGSM), Projected Gradient Descent (PGD), and Carlini-Wagner (C\&W) attacks have been gradually applied in the signal domain \cite{DCariticlerf7,DCariticlerf8,DCariticlerf9,DCariticlerf10,DCariticlerf11}. 
However, these methods often rely on idealized assumptions about real-world communication scenarios, resulting in the generated adversarial examples performing far below expectations by the time they reach the eavesdropper's AMC module \cite{DCariticlerf12}. 
Additionally, the transmitter must ensure that adversarial perturbations remain minimal to avoid degrading communication quality at the receiver \cite{DCariticlerf13}.

Unlike in the computer vision domain, where adversarial example concealment can be intuitively assessed by visually comparing images \cite{DCariticlerf14}, the waveform observed by the signal receiver is influenced by noise and interference during transmission. 
Therefore, directly observing waveform changes cannot accurately quantify the concealment of adversarial perturbations added to the signal. 

\begin{figure}[!t]
\centering
\subfloat[]{\includegraphics[width=8.4cm]{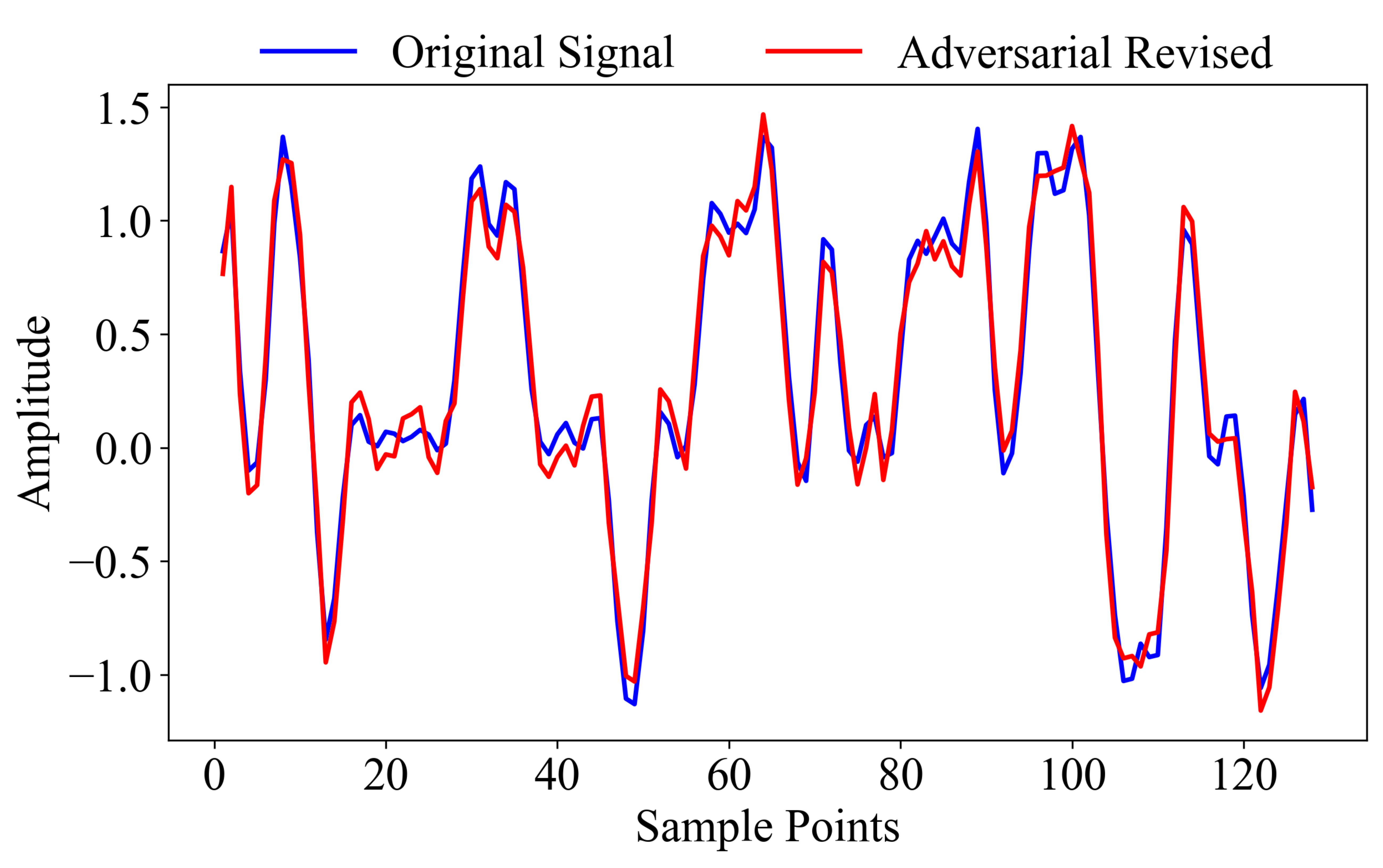}%
\label{fig_first_case}}
\hfil
\subfloat[]{\includegraphics[width=8.4cm]{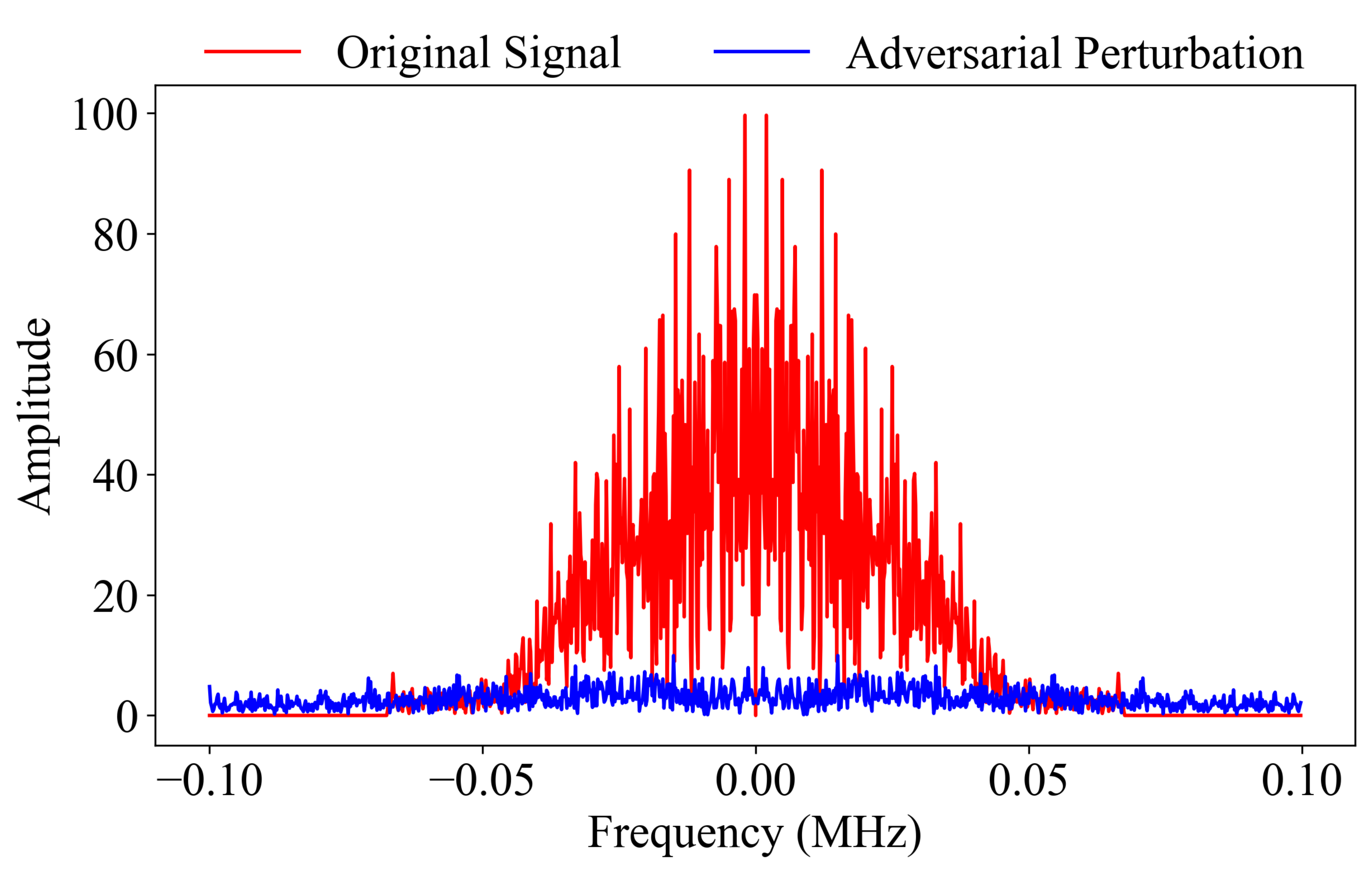}%
\label{fig_second_case}}
\caption{(a) Time-domain waveform of a signal sample before and after adversarial attack (128 sample points are shown for clarity). (b) Spectrum of a signal sample with the added adversarial perturbation.}
\label{fig1}
\end{figure}

The filter, especially the narrowband filter, is one of the most essential and widely used components in signal preprocessing, as it maximizes the retention of useful signal information while eliminating high-frequency noise \cite{DCariticlerf15}. 
However, filters also make it challenging for adversarial perturbations to be effective. 
As illustrated in \figref{fig1}, the perturbations generated by one of the mainstream adversarial example generation algorithms (such as FGSM) exhibit noticeable high-frequency oscillations in the time domain, with energy uniformly distributed across the entire frequency band in the frequency domain. 
To address this, we aim to design a method that constrains the adversarial perturbation energy within the effective frequency band, ensuring its efficacy even when eavesdroppers employ filters for signal preprocessing, while keeping the perturbation energy as minimal as possible.   

For the signal transmitter, we can reasonably infer the filter parameters employed by the eavesdropper. 
In the process of generating adversarial examples, gradients indicate the directions in the input space to which the model is most sensitive \cite{DCariticlerf16}. 
Considering that filter parameters determine the effective range of signal information in the frequency band, we propose introducing frequency-domain constraints into the gradient-based adversarial example generation algorithm to concentrate the perturbation energy within the effective frequency band. 
Inspired by attention mechanisms \cite{DCariticlerf17}, we further apply time-domain constraints to the frequency-constrained adversarial examples to ensure minimal perturbation while overcoming the filter. Additionally, drawing from noise quantification methods in signal processing, we propose a new metric that further enriches the concept of concealment in signal adversarial examples. Experimental results show that the adversarial examples generated by our method not only successfully bypass the eavesdropper's filter and effectively impact the eavesdropper's AMC module, but also modify a smaller portion of the transmitted signal's waveform, appear smoother visually, and exhibit stronger concealment.

\begin{figure}[!t]
\centering
\includegraphics[width=8.4cm]{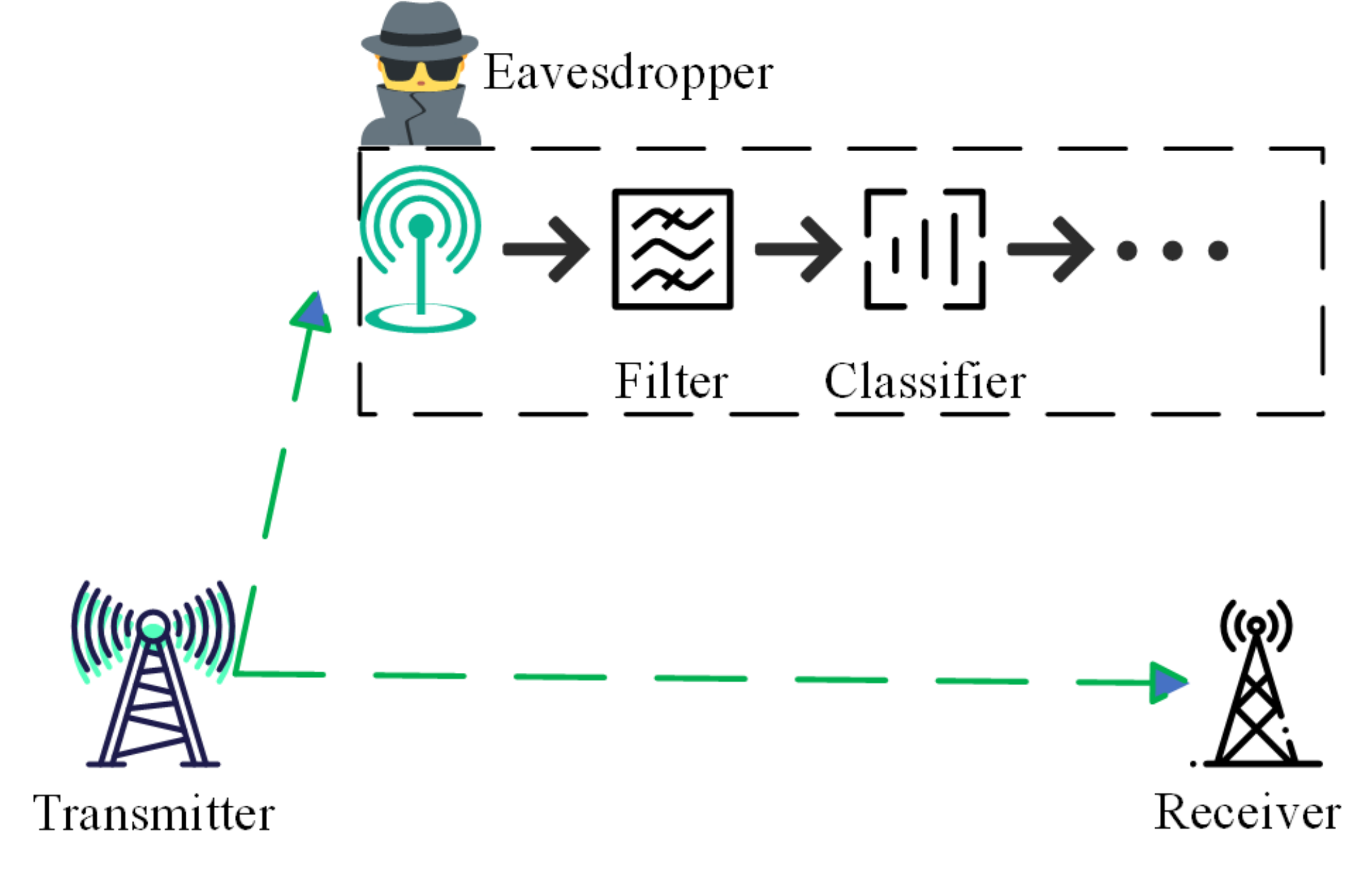}
\caption{System model of a transmitted signal being eavesdropped.}
\label{fig2}
\end{figure}

\section{Proposed Method}
\subsection{Problem Formulation and Review}
As shown in \figref{fig2}, Our work is based on a realistic scenario where an eavesdropper exists within the communication link. The eavesdropper filters the received signal before sending it to the AMC module, followed by other subsequent modules.

For the transmitter, the modulation type label of the transmitted signal $S \in C$  is denoted as $Y \in R$.
The output from the DNN ${f_\theta }$ is denoted as ${Y^p} = {f_\theta }\left( S \right)$.
For a well-trained classifier ${f_\theta }$, the classification result will have a high success rate, yielding ${Y^p} = Y$. 
To prevent the signal from being correctly recognized by the eavesdropper's AMC module, the transmitter introduces a small perturbation to the signal, 
where ${{\left\| \delta  \right\|}_{p}}\le \varepsilon$, and $\varepsilon$ represents a very small positive constant. Thus, the modified signal becomes $S' = S + \delta $, 
Reference \cite{DCariticlerf7} describes the process of finding as a convex optimization problem:
\begin{equation}
\label{deqn_ex1}
\mathrm{max}{L_{{f_\theta }}}\left( {\theta ,S',Y} \right) \quad s.t.{\left\| \delta  \right\|_p} \le \varepsilon \; and \; {Y^p} \ne Y
\end{equation}

Where ${{L}_{{{f}_{\theta }}}}\left( \cdot  \right)$ is the loss function used for training the classifier ${{f}_{\theta }}$. 
Upon the condition $P=\infty $, to solve \eqt{deqn_ex1}, \cite{DCariticlerf7} and \cite{DCariticlerf18} utilize the FGSM, deriving the perturbation expression as: $\delta =\varepsilon \cdot sign\left( {{\nabla }_{S}}{{L}_{{{f}_{\theta }}}}\left( \theta ,S,Y \right) \right)$. 
In this expression, $\nabla $ represents the gradient descent, and $sign\left( \cdot  \right)$ is the sign function.

Furthermore, \cite{DCariticlerf8,DCariticlerf9} utilize the PGD to achieve superior attack performance, meticulously refining the randomly generated initial perturbation through iterative updates and precise projections, and the iterative formula can be described as:
${{{S}'}_{t+1}}=PrD\left( {{{{S}'}}_{t}}+\beta \cdot sign\left( {{\nabla }_{S}}{{L}_{{{f}_{\theta }}}}\left( \theta ,{{{{S}'}}_{t}},Y \right) \right) \right),{{{S}'}_{0}}=S$, 
where $\beta $ signifies the step-size per iteration, and $PrD$ aims to find the closest point in Euclidean space. Reference \cite{DCariticlerf10} utilizes the C\&W method to introduce a regularization term for more precise perturbations, implemented within the PGD framework. 
The iteration formula in this case is described as:

\noindent \resizebox{\columnwidth}{!}{${{{S}'}_{t+1}}=PrD\left( {{{{S}'}}_{t}}+\beta \cdot sign\left( {{\nabla }_{S}}{{L}_{{{f}_{\theta }}}}\left( \theta ,{{{{S}'}}_{t}},Y \right) \right)+\lambda \cdot {{\left\| {{{{S}'}}_{t}}-S \right\|}_{2}} \right)$}

\noindent , ${{{S}'}_{0}}=S$, $\lambda$ with being a constant greater than zero.

In \cite{DCariticlerf11}, a Minimal Power Adversarial Attack (MPA) was introduced, wherein the strength of the adversarial perturbation is defined by adopting the noise intensity measure from the signal domain, namely the perturbation-to-signal ratio (PSR), expressed as: $\mathrm{PSR=10lg}\left( {{{P}_{\mathsf{\delta }}}}/{{{P}_{S}}}\; \right)$. In this expression, ${{P}_{\mathsf{\delta }}}$ corresponds to the energy of the perturbation, while ${{P}_{S}}$ corresponds to that of the signal. \cite{DCariticlerf7,DCariticlerf8,DCariticlerf9,DCariticlerf10,DCariticlerf11} argues that a higher PSR indicates stronger perturbation intensity and greater waveform modification, which in turn leads to poorer concealment.

When the eavesdropper employs a filter with the transfer function $H\left( \cdot  \right)$, the signal reaching the AMC module is $H\left( {{S}'} \right)$. In this scenario, for the perturbation   to continue its effect, \eqt{deqn_ex1} must be reformulated as follows:
\begin{equation}
\label{deqn_ex2}
\mathrm{max}{{L}_{{{f}_{\theta }}}}\left( \theta ,H\left( {{S}'} \right),Y \right) \quad s.t.{{\left\| \delta  \right\|}_{p}}\le \varepsilon \; and \; {{Y}^{p}}\ne Y
\end{equation}

\subsection{Dual-domain Constrained Adversarial Attack}
Narrowband filters are typically designed with linear and time-invariant transfer functions. Consequently, \eqt{deqn_ex2} can be further expressed as follows:
\begin{equation}
\label{deqn_ex3}
\mathrm{max}{{L}_{{{f}_{\theta }}}}\left( \theta ,H\left( S \right)+H\left( \delta  \right),Y \right) \hspace{0.1cm} s.t.{{\left\| \delta  \right\|}_{p}}\le \varepsilon \; and \; {{Y}^{p}}\ne Y
\end{equation}

To solve \eqt{deqn_ex3}, assuming ${{L}_{{{f}_{\theta }}}}\left( \cdot  \right)$ is a mean squared error loss function that is easy to differentiate, the Lagrange multiplier method is employed under condition $P=2$, resulting in the following expression:
\begin{equation}
\label{deqn_ex4}
\varsigma \left( \delta ,\lambda  \right)=\left\| {{f}_{\theta }}\left( H\left( {{S}'} \right) \right)-Y \right\|_{2}^{2}-\lambda \left( \left\| \delta  \right\|_{2}^{2}-\varepsilon  \right), \; \lambda>0
\end{equation}

Under the satisfaction of the Karush-Kuhn-Tucker (KKT) conditions, the expression for the perturbation is obtained as follows:
$\delta  =  - {\left( {{H^T}H + \lambda } \right)^{ - 1}}{H^T}\left( {{f_\theta }\left( {H\left( S \right)} \right) - Y} \right)$.

Inspired by \cite{DCariticlerf18}, adversarial examples can be viewed as features. Coupled with the solutions from the formulation \eqt{deqn_ex4}, We consider the process of solving equation \eqt{deqn_ex3} as limiting the iterative direction by incorporating filter parameters during gradient calculation. Following the PGD algorithm workflow, this approach can be extended to cases where the loss function is difficult to differentiate (e.g., cross-entropy). The detailed process is shown in Algorithm \ref{alg1}:

  \begin{algorithm}
    \caption{Adversarial Example with Frequency Domain Constraints}
    \label{alg1}
    \begin{algorithmic}[1]
        \renewcommand{\algorithmicrequire}{\textbf{Input:}} 
        \REQUIRE Original signal example $S$; Modulation ground-truth label $Y$; Loss function ${{L}_{{{f}_{\theta }}}}$; The filter coefficients $H$.
        \renewcommand{\algorithmicrequire}{\textbf{Initialization:}}
        \REQUIRE The perturbation size $\varepsilon $; The iteration $N$;\\ ${{{S}'}_{0}}=H\cdot S$; The Step-size $\beta =\varepsilon /N$.
        \renewcommand{\algorithmicensure}{\textbf{Output:}}
        \ENSURE An adversarial example ${S}'$, where ${{\left\| {S}'-S \right\|}_{\infty }}\le \varepsilon \text{}$.
        \renewcommand{\algorithmicensure}{\textbf{Iterations:}}
        \ENSURE for $t=0:N-1$:
        \STATE \qquad Calculate the gradient of ${{L}_{{{f}_{\theta }}}}$: ${{\nabla }_{{{{{S}'}}_{t}}}}{{L}_{{{f}_{\theta }}}}$;
        \STATE \qquad Constrain the gradient:$H\cdot {{\nabla }_{{{{{S}'}}_{t}}}}{{L}_{{{f}_{\theta }}}}$;
        \STATE \qquad Update ${{{S}'}_{t+1}}={{{S}'}_{t}}+\beta sign\left( H\cdot {{\nabla }_{{{{{S}'}}_{t}}}}{{L}_{{{f}_{\theta }}}} \right)$;
        \STATE \quad end for 
        \STATE \quad return ${S}'={{{S}'}_{N}}$
    \end{algorithmic}
  \end{algorithm}

Algorithm \ref{alg1} applies frequency-domain constraints to the perturbation generation process based on the band-pass filter's characteristics in filtering signal frequencies. To further ensure communication quality, it is crucial to keep the perturbations as minimal as possible. Inspired by attention mechanisms \cite{DCariticlerf17}, Gradient-weighted Class Activation Mapping (Grad-CAM) is employed to derive the heatmap function $H_{_{map}}^{ori}$ of the classifier ${{f}_{\theta }}$. This heatmap is then up-sampled to match the original signal dimensions using bilinear interpolation. Subsequently, the data is normalized to the range [0,1], resulting in $H_{_{map}}^{fin}$. A threshold of 0.6 is set to identify the regions where the signal receives high attention from the model, resulting in the mask function ${{M}_{ask}}$. The time-domain constraint on the perturbation is then given by the following expression:
\begin{equation}
\label{deqn_ex5}
\begin{aligned}
  &                \delta ={{\delta }_{F}}\cdot {{M}_{ask}},\quad {{\delta }_{F}}={S}'-S \\ 
 & {{M}_{ask}}\left( S \right)=\left\{ \begin{matrix}
   1,\quad if \quad H_{_{map}}^{fin}\left( {{S}_{k}} \right)\ge 0.6  \\
   0,\quad if \quad H_{_{map}}^{fin}\left( {{S}_{k}} \right)<0.6  \\
\end{matrix} \right. \\ 
\end{aligned}
\end{equation}

Where ${{\delta }_{F}}$ denotes the perturbation derived from Algorithm \ref{alg1}, and ${{S}_{k}},k=1\cdots K$ represents the $k$-th point of the signal $S$, which has a total length $K$. 

By applying Algorithm \ref{alg1} and \eqt{deqn_ex5}, we have developed an adversarial perturbation that is simultaneously constrained in the time and frequency domains. This method is named by us the Dual-domain Constrained Adversarial Attack.

In this work, we posit that the portions of perturbations that can be filtered out by the filter cannot be used to quantify the concealment of the perturbation. Therefore, by integrating domain knowledge of the signal, we employ the Narrowband Perturbation-to-Signal Ratio (NB-PSR) for evaluation. The expression is as follows:
\begin{equation}
\label{deqn_ex6}
\begin{aligned}
  \mathrm{NB-PSR}=10\lg \left( \frac{{{P}_{H\left( \mathbf{\delta } \right)}}}{{{P}_{H\left( S \right)}}} \right)
\end{aligned}
\end{equation}

\begin{figure*}[!ht]
\centering
\subfloat[FGSM]{\includegraphics[width=5.6cm]{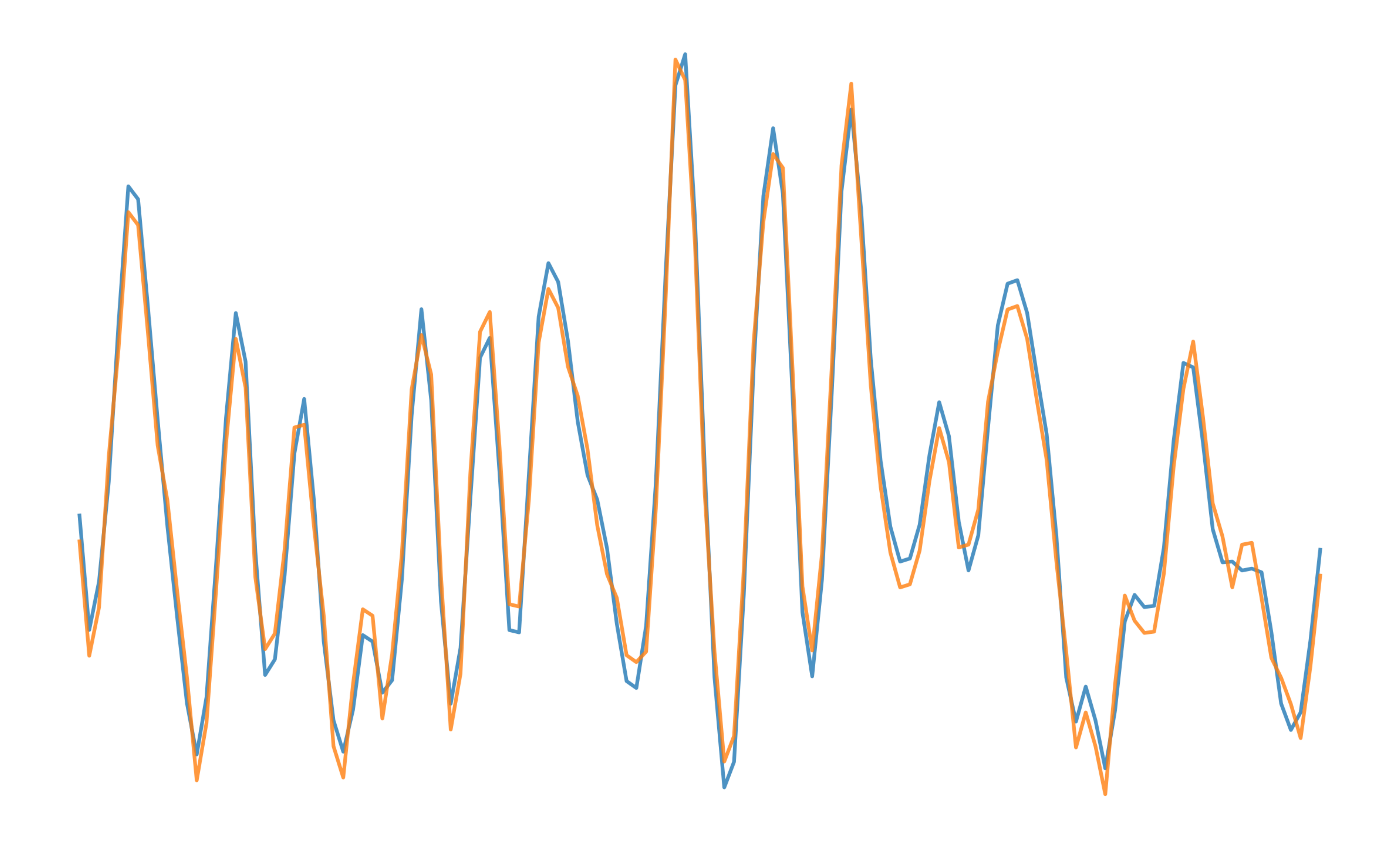}%
\label{fig3a}} \hspace{0.5cm}
\subfloat[PGD]{\includegraphics[width=5.6cm]{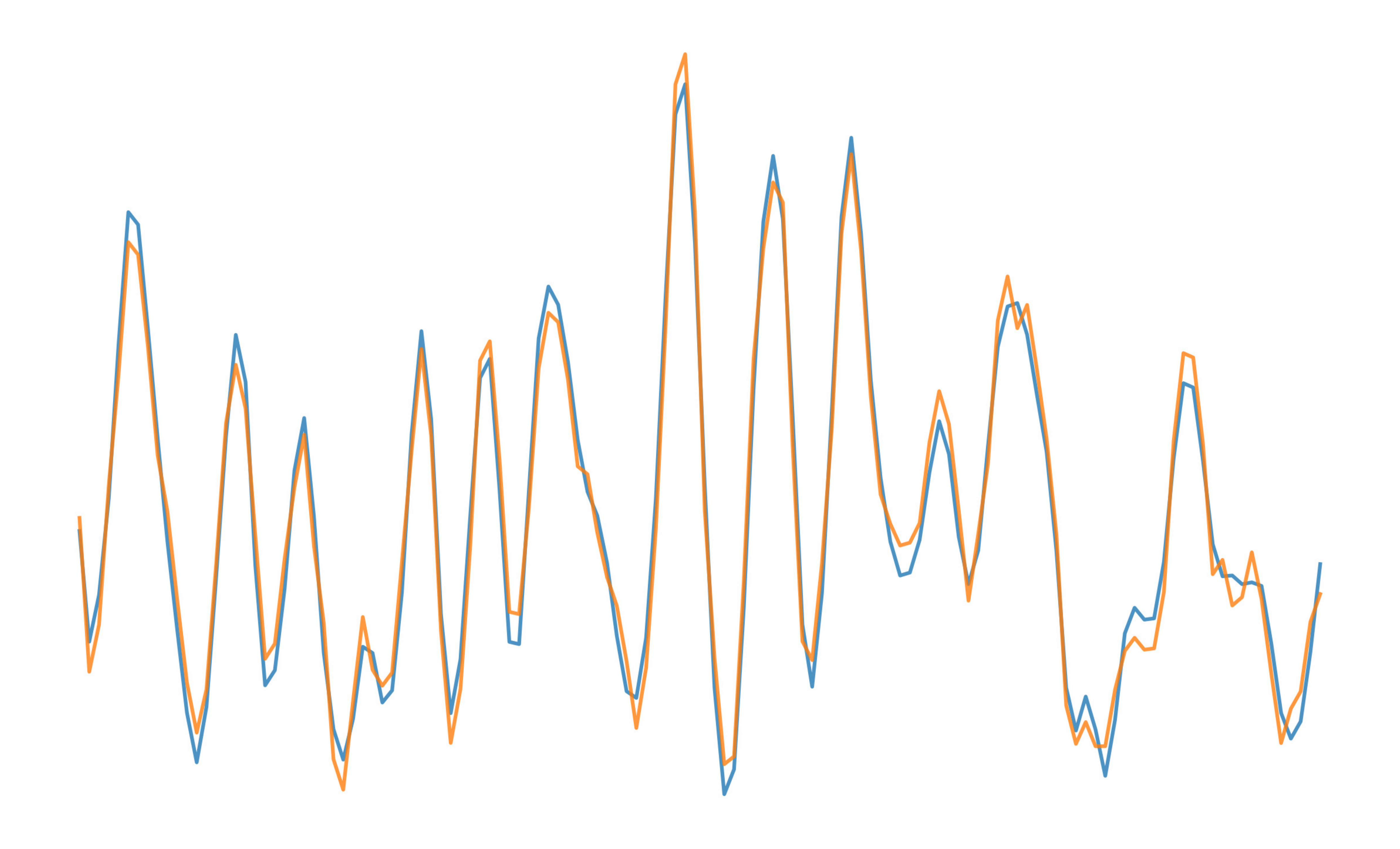}%
\label{fig3b}} \hspace{0.5cm}
\subfloat[C\&W]{\includegraphics[width=5.6cm]{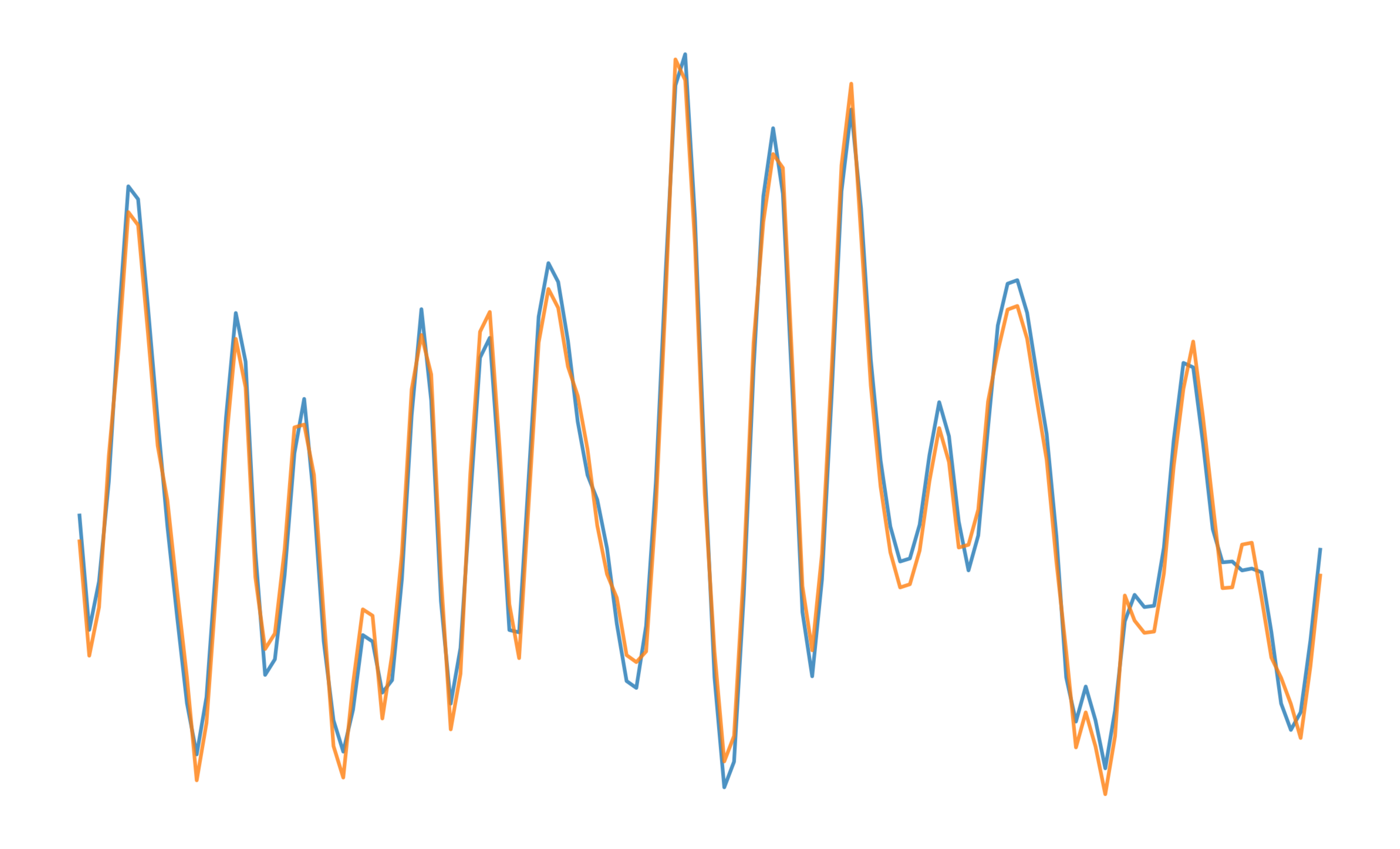}%
\label{fig3c}} \\[1ex]
\subfloat[MPA]{\includegraphics[width=5.6cm]{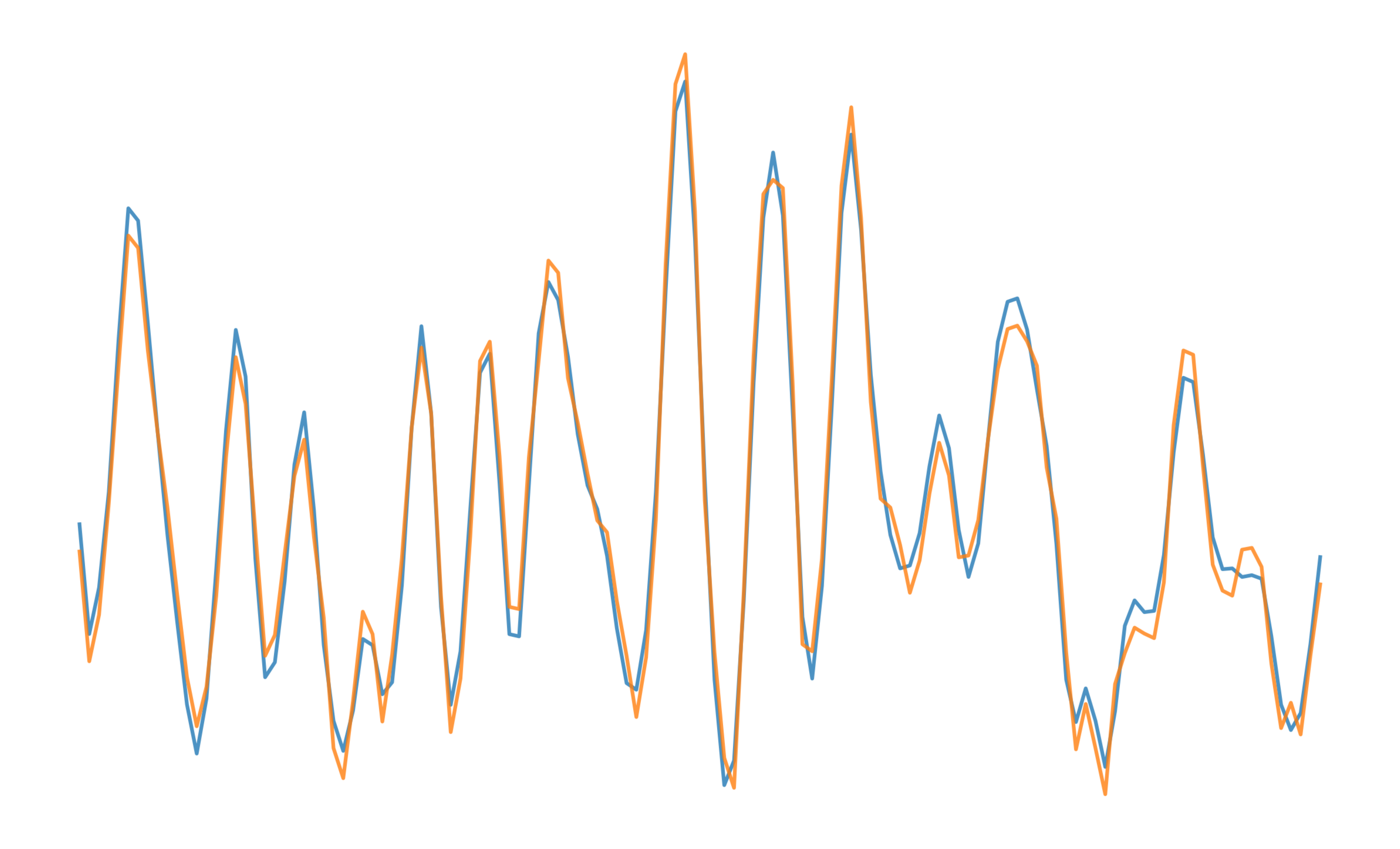}%
\label{fig3d}} \hspace{0.5cm}
\subfloat[DC-PDG]{\includegraphics[width=5.6cm]{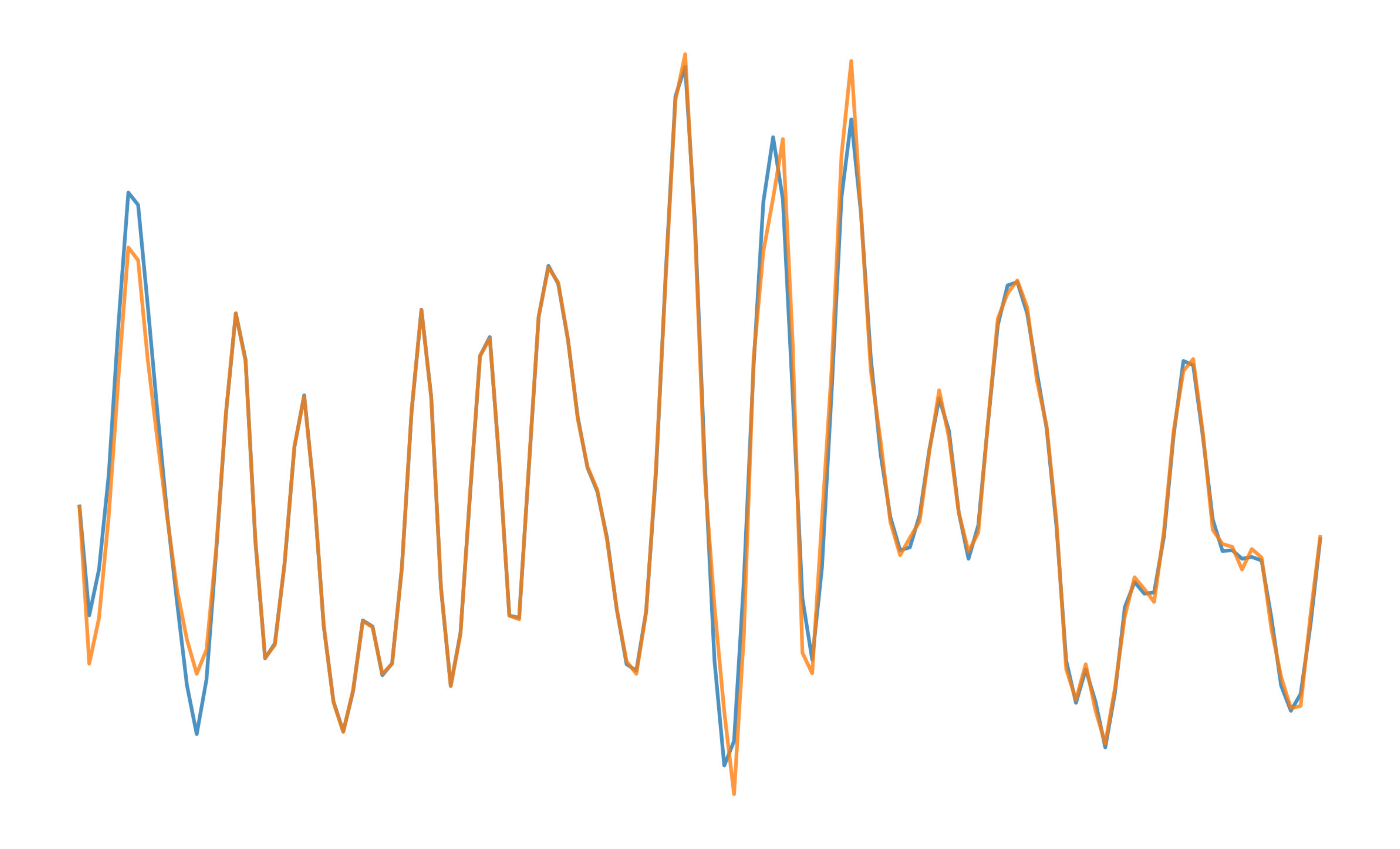}%
\label{fig3e}}
\caption{Waveform comparison of a 16QAM modulated signal with an SNR of 18 dB before and after perturbation by five different adversarial attack methods.(128 sample points are shown for clarity; the blue line indicates the original signal, and the orange line indicates the adversarial example)}
\label{fig3}
\end{figure*}

\begin{figure*}[!ht]
\centering
\subfloat[FGSM]{\includegraphics[width=5.6cm]{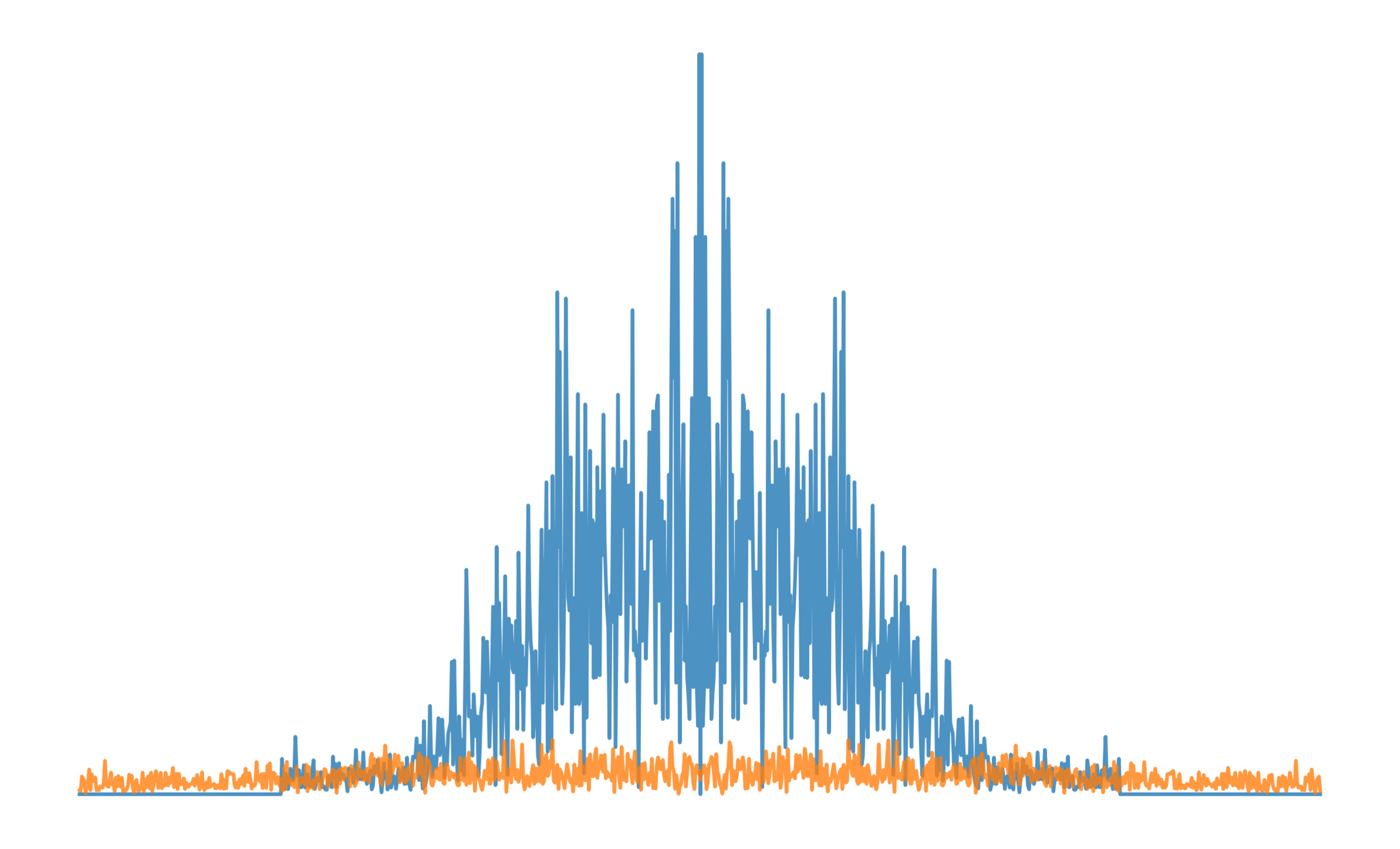}%
\label{fig4a}} \hspace{0.5cm}
\subfloat[PGD]{\includegraphics[width=5.6cm]{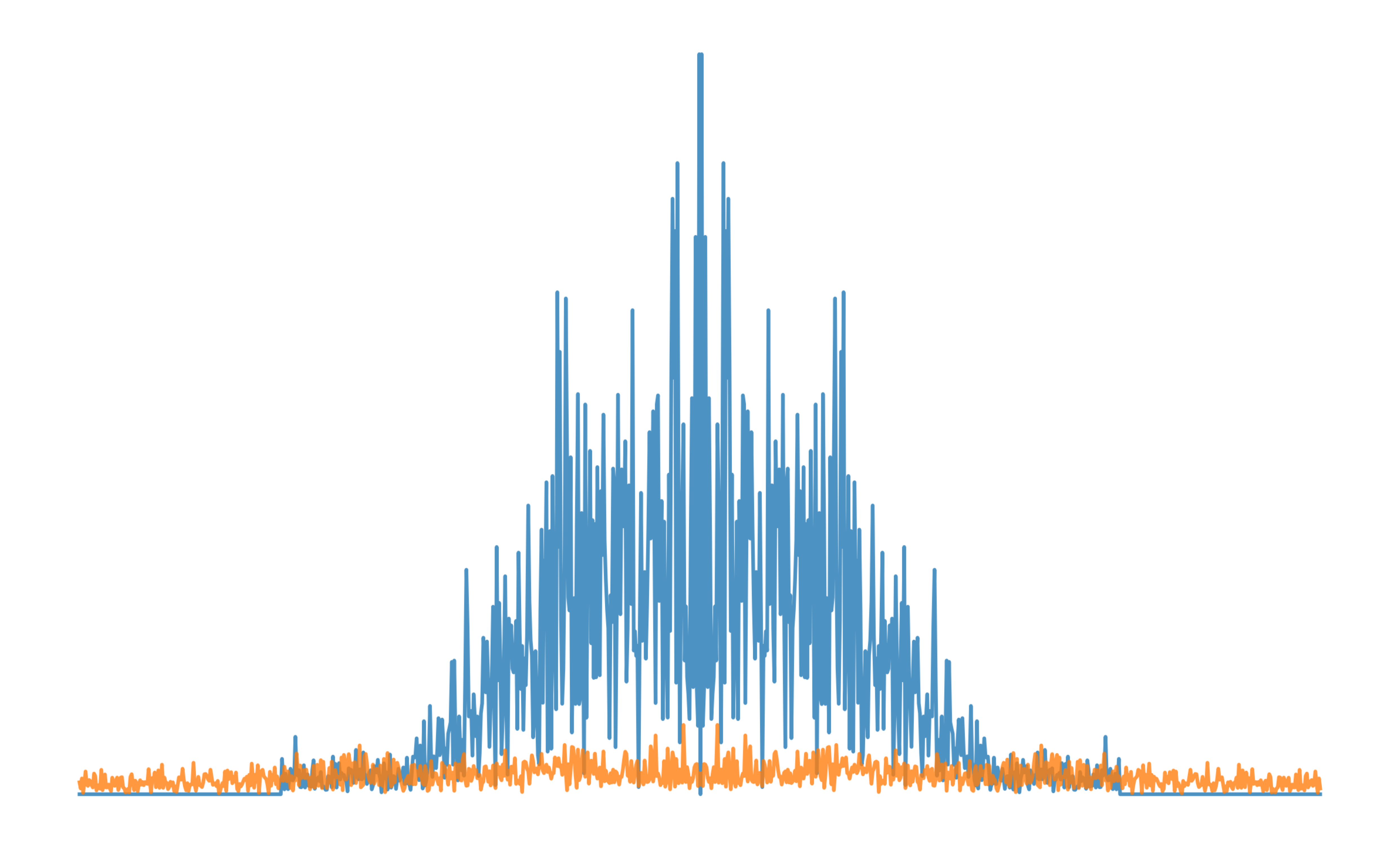}%
\label{fig4b}} \hspace{0.5cm}
\subfloat[C\&W]{\includegraphics[width=5.6cm]{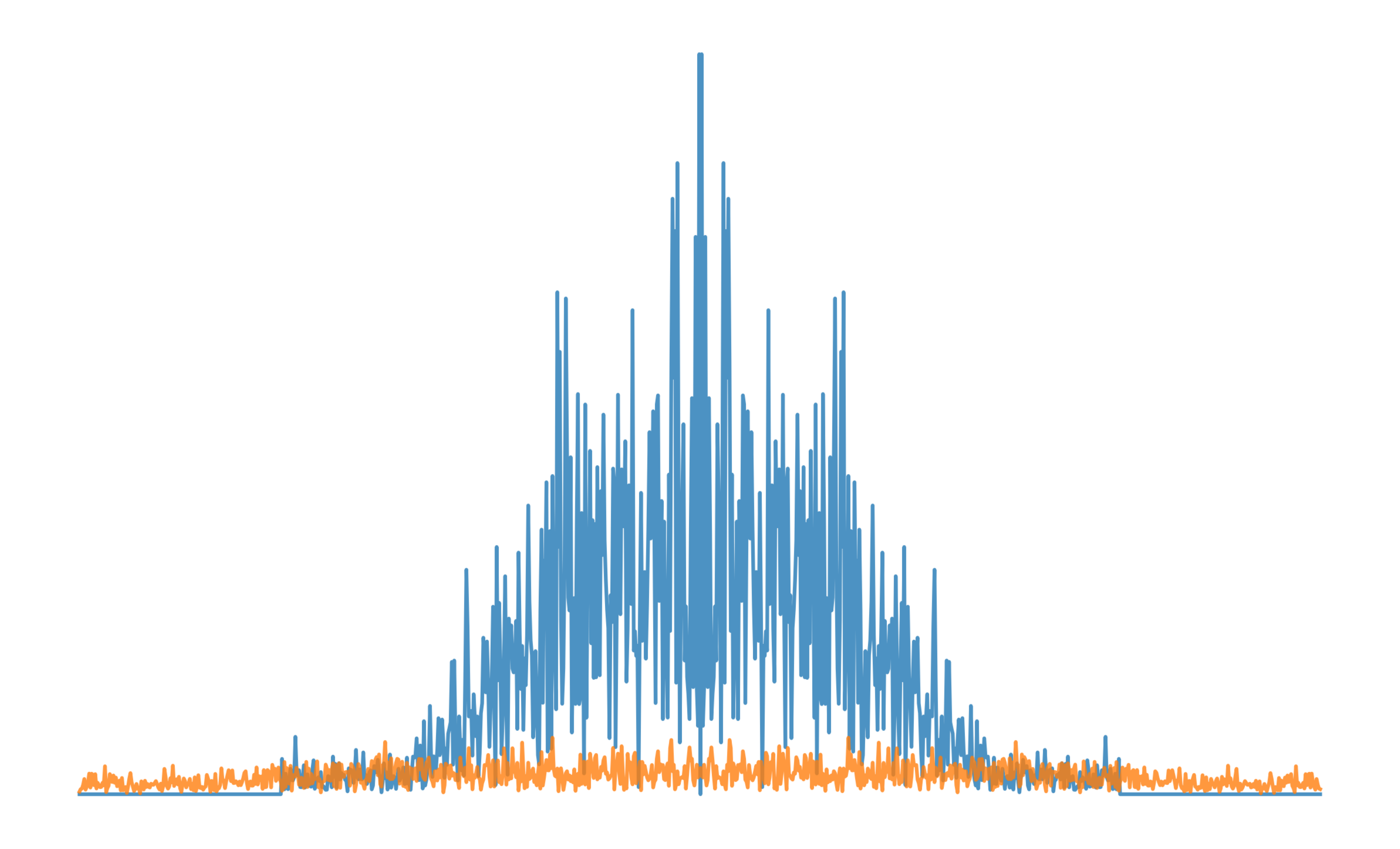}%
\label{fig4c}} \\[1ex]
\subfloat[MPA]{\includegraphics[width=5.6cm]{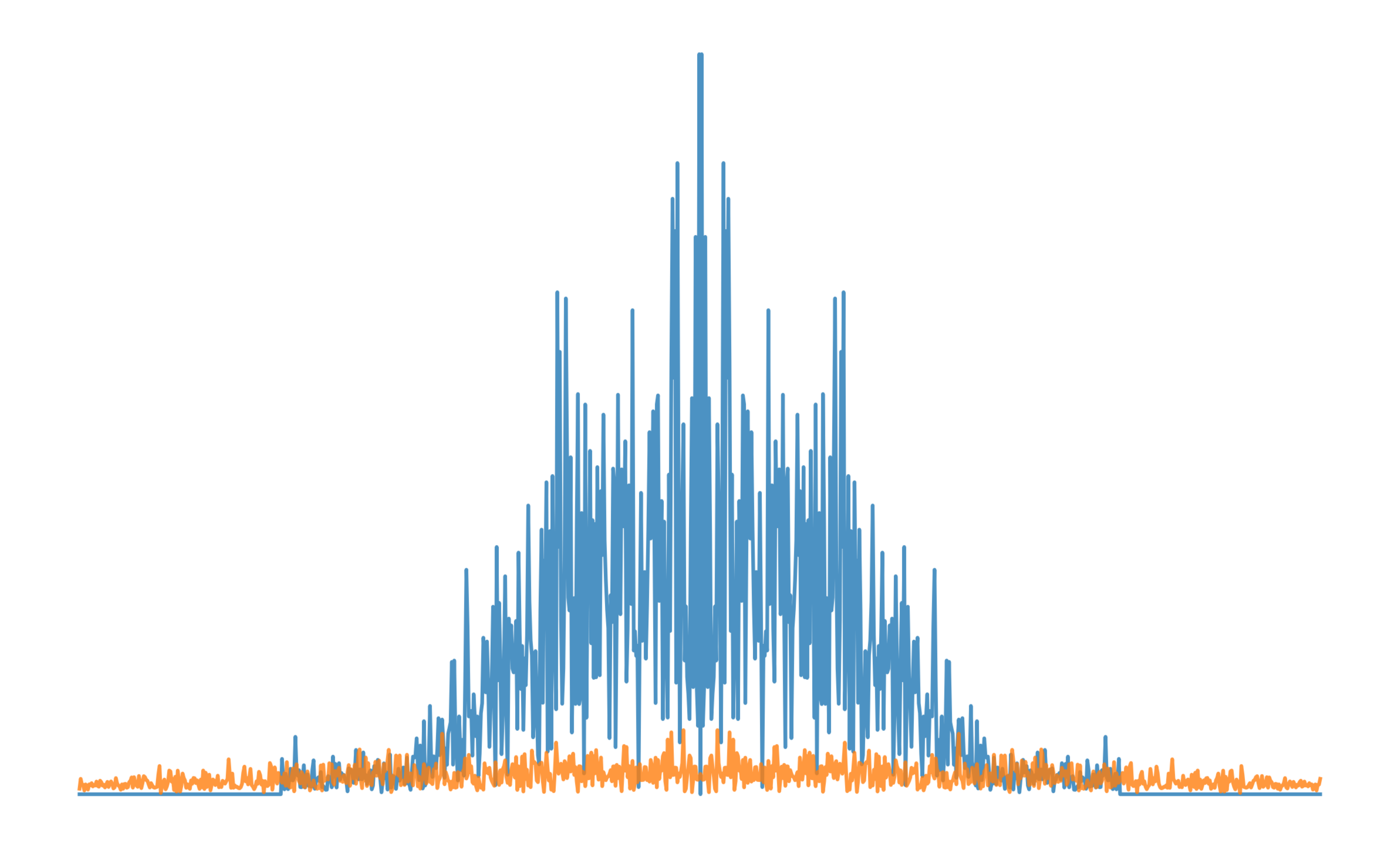}%
\label{fig4d}} \hspace{0.5cm}
\subfloat[DC-PDG]{\includegraphics[width=5.6cm]{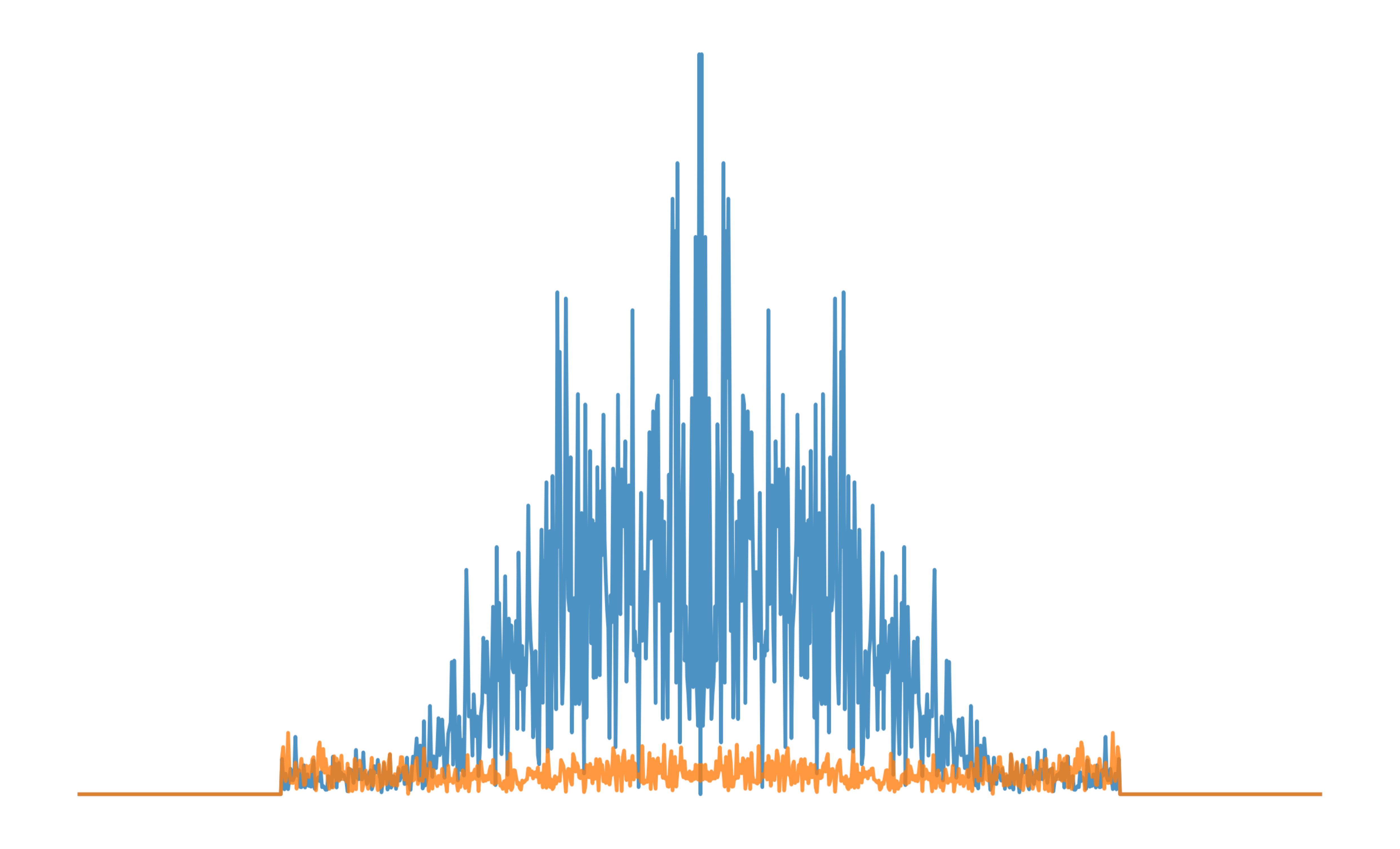}%
\label{fig4e}}
\caption{Spectrum of the perturbation generated by five different adversarial attack methods on a 16QAM modulated signal with an SNR of 18 dB.}
\label{fig4}
\end{figure*}

\section{Experiments}
To verify the performance of the proposed algorithm, corresponding experiments were conducted, and the modified signals were visualized. The dataset used in this study is the open-source radio dataset HisarMod2019.1 \cite{DCariticlerf19}.

\subsection{Experimental Setup}
The HisarMod2019.1 dataset encompasses 26 commonly used modulation types, namely BPSK, QPSK, 8PSK, 16PSK, 32PSK, 64PSK, 4QAM, 8QAM, 16QAM, 32QAM, 64QAM, 128QAM, 256QAM, 2FSK, 4FSK, 8FSK, 16FSK, 4PAM, 8PAM, 16PAM, AM-DSB, AM-DSB-SC, AM-USB, AM-LSB, FM, and PM. 
Each modulation type is sampled across SNR values ranging from -20 dB to 18 dB at 2 dB increments, with 1,500 samples collected for every SNR level. 
Each signal sample consists of 1024 in-phase and quadrature (I/Q) data points, with an oversampling rate of 2. Additionally, a raised-cosine pulse-shaping filter with a roll-off factor of 0.35 is employed.

Regarding the transmission environment, five distinct wireless channel models are included: additive white Gaussian noise (AWGN), static channel, Rayleigh fading, Rician fading (k=3), and Nakagami-m fading (m=2). 
After capturing a single signal sample, the eavesdropper uses a pre-trained AMC model to identify the modulation type. 
Our work only uses samples with a SNR$\geq$10dB, and these samples are divided into training, validation, and test sets in a 8:2:5 ratio.

The target model is trained using the Adam optimizer and cross-entropy loss over a maximum of 200 epochs. An early-stopping mechanism is applied with a patience of 30. Through this procedure, a ResNet18-based classifier is well trained to identify the modulation of the received signals.

To evaluate the effectiveness of the proposed algorithm, we adopted FGSM, PGD, C\&W and MPA as benchmark methods, comparing them against the Dual-domain Constrained PGD (DC-PGD) proposed in this study. 
The perturbation strength was set to PSR = -20 dB, the number of iterations was set to 10, and the baseband sampling rate was set to 200 kHz. Signals with an SNR of 18 dB from the test sets were used for evaluation. 
To ensure the integrity of the received signal, the eavesdropper uses a filter with at least 135 kHz bandwidth. As the signal transmitter, we can easily infer the signal's bandwidth. 
The effectiveness of the adversarial example in protecting the original signal is shown by its ability to attack the eavesdropper's classifier, measured by the Fool Rate (FR):
\begin{equation}
\label{deqn_ex7}
\begin{aligned}
  \mathrm{FR}=\frac{1}{n}\sum\limits_{i=1}^{n}{\ell \left( {{f}_{\theta }}\left( {{S}'} \right)\ne {{f}_{\theta }}\left( S \right) \right)\text{ }},f\left( S \right)=Y
\end{aligned}
\end{equation}

$\ell \left( \centerdot  \right)$ denotes the indicator function, which takes the value 1 if the condition is met and 0 otherwise. The filter based on the Fast Fourier Transform (FFT) is used in the experiment.

\subsection{Attack Performance}
\subfigrefcon{fig3}{fig3a}{fig3e} display the waveform of 16QAM signals at an SNR of 18 dB, both before and after modification by five different methods, the blue line represents the original signal, while the orange line shows the signal modified by adversarial perturbations.
\subfigrefcon{fig4}{fig4a}{fig4e}, the orange line corresponds to the spectrum of the adversarial perturbation, indicating that the perturbation's energy is confined within the signal's frequency band. 
\tble{tab1} evaluates the performance of five methods using five metrics with a 135 kHz bandwidth FFT filter. 
\tble{tab2} lists the average FR of DC-PDG modified signals after passing through four different filters (FFT, Chebyshev I, Chebyshev II, and Butterworth) with bandwidths ranging from 135 kHz to 155 kHz.

\begin{table}[ht]
\centering
\caption{Results Of The Fr By Five Attack Approaches}
\begin{tabular}{|c|c|c|c|c|c|}
  \hline
  & \multicolumn{1}{c|}{\textbf{FR}} & \multicolumn{1}{c|}{\textbf{FR}} &  &  &  \\ 
 \textbf{Method} & \textbf{(without} & \textbf{(passing} & \textbf{$\Delta$ FR} & \textbf{NB-PSR} & \textbf{$\Delta$ PSR} \\ 
  & \textbf{filter)} & \textbf{filter)} &  &  &  \\ \hline
FGSM & 91.88\% & 41.08\% & -50.80\% & 21.59dB & -1.59dB \\ \hline
PGD & 97.65\% & 44.62\% & -53.03\% & 21.38dB & -1.38dB \\ \hline
C\&W & 97.01\% & 45.26\% & -51.75\% & 21.25dB & -1.25dB \\ \hline
MPA & 92.55\% & 44.07\% & -48.48\% & 21.47dB & -1.47dB \\ \hline
DC-PGD & 92.84\% & 90.17\% & -2.67\% & 20.27dB & -0.27dB \\ \hline
\end{tabular}
\label{tab1}
\end{table}

\begin{table}[ht]
\centering
\caption{The Fr Of DC-PGD Under Different Filter Setting}
\begin{tabular}{|c|c|c|c|c|c|}
\hline
    & \textbf{135kHz} & \textbf{145kHz} & \textbf{155kHz} \\ \hline
FFT & 90.17\% & 89.55\% & 90.02\% \\ \hline
Chebyshev I & 89.73\% & 87.90\% & 88.29\% \\ \hline
Chebyshev II & 88.11\% & 88.92\% & 88.38\% \\ \hline
Butterworth & 88.25\% & 89.18\% & 88.89\% \\ \hline
\end{tabular}
\label{tab2}
\end{table}

\section{Discussion}
As shown in \subfigrefcon{fig3}{fig3a}{fig3d}, the waveforms of the signals modified by FGSM, PGD, C\&W, and MPA exhibit noticeable abrupt jitter, corresponding to high-frequency components of the signals. 
In contrast, the left portion of \subfigref{fig3}{fig3e} shows that the DC-PGD-modified waveform introduces smoother perturbations that target the low-frequency components. Due to the attention mechanism, the time-domain region of the modified waveform is effectively constrained. 
As shown in \subfigrefcon{fig4}{fig4a}{fig4d}, the perturbation energy is spread across the entire frequency band, whereas in \subfigref{fig4}{fig4e}, the DC-PGD-generated adversarial perturbations are limited to the signal's bandwidth, thus achieving frequency-domain constraints.

\tble{tab1} shows that adversarial examples generated by five methods for a signal with an SNR of 18 dB achieve more than 90\% FR before filtering. 
However, after passing through an FFT filter with a 135 kHz bandwidth, the attack performance of FGSM, PGD, C\&W, and MPA methods experiences significant degradation ($\Delta$FR = -50.80\%, -53.03\%, -51.75\%, -48.48\%), while the proposed DC-PGD method demonstrates much more stable performance, with a $\Delta$FR of -2.67\%. 
Furthermore, the NB-PSR results provide additional confirmation of the frequency-domain energy constraint imposed by our method. Compared to the other four approaches, the perturbations generated by DC-PGD exhibit the lowest $\Delta$PSR (-0.27 dB), maintaining the ability to overcome eavesdropper filters. Thus, NB-PSR serves as a measure of the perturbation's concealment relative to the eavesdropper's filter. Perturbations generated by different method with the same PSR, a higher NB-PSR indicates greater energy evading the filter, reflecting stronger concealment.

\tble{tab2} shows that DC-PGD maintains high FR performance (above 85\%) across different filter configurations. Whether or not the eavesdropper filter's bandwidth matches our signal, DC-PGD performs robustly, adapting to various filter types and parameters.

\section{Conclusion}
We consider a more realistic wireless communication scenario where an eavesdropper filters the signal before performing AMC, rendering existing attacks ineffective. 
Our Dual-domain Constrained adversarial attack framework targets the frequency domain subspace. 
Experimental results show that our approach outperforms others, maintaining attack performance post-filtering by the eavesdropper across various filter settings. 
Additionally, it results in smaller waveform modification regions, smoother perturbations, and better concealment visually. 
Future work will explore the black-box scenarios with unknown eavesdropper AMC models and further study the impact of channel fading on adversarial perturbation, bringing our attacks closer to real-world.
\balance
\begingroup
\bibliographystyle{IEEEtran}
\bibliography{IEEEabrv,main} 
\endgroup

\end{document}